\documentclass[journal=jacsat,manuscript=perspective]{achemso}
\UseRawInputEncoding
\usepackage[version=3]{mhchem} 
\usepackage{graphicx}
\usepackage{dcolumn}
\usepackage{bm}
\usepackage{amsfonts, amsmath, amsthm, amssymb}
\usepackage{hyperref}
\usepackage[mathlines]{lineno}

\usepackage[dvipsnames]{xcolor}
\author{Zhi-He Hao}
\affiliation{CAS Key Laboratory of Quantum Information, University of Science and Technology of China, Hefei, Anhui 230026, China}
\affiliation{
Anhui Province Key Laboratory of Quantum Network, University of Science and Technology of China, Hefei, Anhui 230026, China.}
\affiliation{CAS Center For Excellence in Quantum Information and Quantum Physics,University of Science and Technology of China, Hefei, Anhui 230026, China}
\author{Zhen-Xuan He}
\affiliation{CAS Key Laboratory of Quantum Information, University of Science and Technology of China, Hefei, Anhui 230026, China}
\affiliation{
Anhui Province Key Laboratory of Quantum Network, University of Science and Technology of China, Hefei, Anhui 230026, China.}
\affiliation{CAS Center For Excellence in Quantum Information and Quantum Physics,University of Science and Technology of China, Hefei, Anhui 230026, China}
\affiliation{Hefei National Laboratory, University of Science and Technology of China, Hefei 230088, China}
\author{Jovan Maksimovic}
\affiliation{Optical Sciences Center and ARC Training Centre in Surface Engineering for Advanced Materials (SEAM), Swinburne University of Technology, John Street, Hawthorn, Victoria 3122, Australia. }
\author{Tomas Katkus}
\affiliation{Optical Sciences Center and ARC Training Centre in Surface Engineering for Advanced Materials (SEAM), Swinburne University of Technology, John Street, Hawthorn, Victoria 3122, Australia. }
\author{Jin-Shi Xu}
\affiliation{CAS Key Laboratory of Quantum Information, University of Science and Technology of China, Hefei, Anhui 230026, China}
\affiliation{
Anhui Province Key Laboratory of Quantum Network, University of Science and Technology of China, Hefei, Anhui 230026, China.}
\affiliation{CAS Center For Excellence in Quantum Information and Quantum Physics,University of Science and Technology of China, Hefei, Anhui 230026, China}
\affiliation{Hefei National Laboratory, University of Science and Technology of China, Hefei 230088, China}
\email{jsxu@ustc.edu.cn}
\author{Saulius Juodkazis}
\affiliation{Optical Sciences Center and ARC Training Centre in Surface Engineering for Advanced Materials (SEAM), Swinburne University of Technology, John Street, Hawthorn, Victoria 3122, Australia. }
\affiliation{Laser Research Center, Physics Faculty, Vilnius University, Saul\.{e}tekio Ave. 10, 10223 Vilnius, Lithuania}
\author{Chuan-Feng Li} 
\affiliation{CAS Key Laboratory of Quantum Information, University of Science and Technology of China, Hefei, Anhui 230026, China}
\affiliation{
Anhui Province Key Laboratory of Quantum Network, University of Science and Technology of China, Hefei, Anhui 230026, China.}
\affiliation{CAS Center For Excellence in Quantum Information and Quantum Physics,University of Science and Technology of China, Hefei, Anhui 230026, China}
\affiliation{Hefei National Laboratory, University of Science and Technology of China, Hefei 230088, China}
\email{cfli@ustc.edu.cn }
\author{Guang-Can Guo}
\affiliation{CAS Key Laboratory of Quantum Information, University of Science and Technology of China, Hefei, Anhui 230026, China}
\affiliation{
Anhui Province Key Laboratory of Quantum Network, University of Science and Technology of China, Hefei, Anhui 230026, China.}
\affiliation{CAS Center For Excellence in Quantum Information and Quantum Physics,University of Science and Technology of China, Hefei, Anhui 230026, China}
\affiliation{Hefei National Laboratory, University of Science and Technology of China, Hefei 230088, China}
\author{Stefania Castelletto}
\affiliation{%
School of  Engineering, RMIT University, Melbourne, Victoria 3001, Australia.
}%
\email{stefania.castelletto@rmit.edu.au}

\title{Laser writing and spin control of near infrared emitters in silicon carbide}
\keywords{Silicon carbide, Photoluminescence, point defects in the bandgap, laser writing, spin coherence, divacancy}
\begin{document}
\begin{abstract}
Near infrared emission in silicon carbide is relevant for quantum technology specifically single photon emission and spin qubits for integrated quantum photonics, quantum communication and quantum sensing. In this paper we study the fluorescence emission of direct femtosecond laser written array of color centres in silicon carbide  followed by thermal annealing. We show that in high energy laser writing pulses regions a near telecom O-band ensemble fluorescence emission is observed after thermal annealing and it is tentatively attributed to the nitrogen vacancy centre in silicon carbide. Further in the low energy laser irradiation spots after annealing, we fabricated few divacancy, PL5 and PL6 types and demonstrate their optical spin read-out, and coherent spin manipulation (Rabi and Ramsey oscillations and spin echo). We  show that direct laser writing and thermal annealing can yield bright near telecom emission and preserve the spin coherence time of divacancy at room temperature. 
\end{abstract}
\maketitle

\section{Introduction}
Silicon Carbide (SiC) is a third generation semiconductor used in the high-power electronic industry, for high temperature\cite{casady1996status} and high-frequency applications. In its more commonly deployed hexagonal polytype (4H-SiC), it is currently commercially available in eight inches wafer scale with the desired intrinsic epilayers purity, and n and p-type doping control. SiC is a CMOS compatible semiconductor with linear and nonlinear optical properties, ideal for integrated photonic applications\cite{lukin2020integrated,castelletto2022silicon,bader2024analysis}. Recent studies have demonstrated that SiC paramagnetic color centres electron spin can serve as qubit\cite{wolfowicz2021quantum}, can be initialized, read out in a single shot and coherently controlled with coherence time up to five second \cite{anderson2021five}.  A quantum coherent spin-photon interface\cite{Nagy2019,Son2020perspective,morioka2020spin} can be created based on SiC spin selective optical transitions to achieve spin-photon entanglement\cite{Fang2024spin_photon} and enable connection between the quantum memory of the solid-state qubit, coupled nuclear spins\cite{bourassa2020entanglement}, and the flying qubit encoded into single photon. These achievements have established SiC as a platform for quantum communication, quantum computing architectures and quantum sensing.  The integration of these spin-photon qubits in conventional devices or tailored devices can be pursued in this semiconductor to scale quantum technologies engineering into devices with enhanced sensitivities. 
Specifically, relevant qubits in SiC have been associated to spin S=1 defects, such as the divacancy centres ($\rm V_{Si}\rm V_{C}^0$)  \cite{koehl_room_2011, falk_polytype_2013, christle_isolated_2015, klimov_quantum_2015, zwier_all-optical_2015,zhou2021experimental,anderson2021five,li2022room}, consisting of neighboring C and Si vacancies. There are several emissions in 4H-SiC that are attributed to divacancies as determined by their photoluminescence (PL), optically detected magnetic resonance (ODMR) and spin control studies, known as PL1-PL7, with  PL1 (1132 nm), PL2 (1131 nm), and PL6 (1038 nm) along the $c$-axis, PL3 (1108 nm), PL4 (1078 nm), PL5 (1043 nm), and PL7 (unknown PL) along (0001) basal plane\cite{koehl_room_2011, christle_isolated_2015}. Details of the properties of divacancies in bulk 4H-SiC and their applications for quantum sensing of magnetic field and temperature were recently reviewed \cite{Castelletto_2024}.
Divancacies are generated by thermal annealing from the silicon monovacancy ($\rm V_{Si}^-$) in 4H-SiC \cite{Widmann2014a, babin2021nanofabricated}, which is also an excellent spin-photon interface with S=3/2 in the spectral region of 920 nm. Selection from one to another emitter is primarily achieved using thermal annealing, as for instance the $\rm V_{Si}^-$ anneals out above 600$^\circ$ C and and divacancy emerges at 900 $^\circ$ C annealing.
A telecom O-band emission in 4H-SiC is attributed to the nitrogen vacancy ($\rm N_{C} V_{Si}^-$) centres \cite{von_bardeleben_identification_2015, von_bardeleben_nv_2016, zargaleh_electron_2018, mu_coherent_2020, wang_coherent_2020}. The ($\rm N_{C} V_{Si}^-$) in 4H-SiC also possess four ZPLs depending on the location in the crystal, namely 1241 nm (kk) and 1223.3 nm (hh) on-axis and 1180 nm (kh) and 1242 nm (hk) off-axis \cite{wang_coherent_2020}. $\rm N_{C} V_{Si}^-$ is created after nitrogen ions implantation in intrinsic 4H-SiC epilayers and thermal annealing at around 1050$^\circ$ C \cite{wang_coherent_2020} or it can be created in ensemble by hydrogen ions implantation in N-doped 4H-SiC \cite{CastellettoDeterministic2019, mu_coherent_2020}. Room temperature spin control of nitrogen ions implanted $\rm N_{C} V_{Si}^-$ with coherence time of 17.1 $\mu$s and single photon emission have been demonstrated at room temperature\cite{wang_bright_2018}.

Currently the fabrication of these color centres is based on electron irradiation, ions implantation and focused ions beams \cite{bader2024analysis,Luofabrication2023}. Helium focused ions beam (FIB) can be used to fabricate these emitters in array without masks \cite{Zhen-Xuan2023}, which is very desirable for placing the qubits into photonic nanostructures, however the control of their formation at depth is limited to 179 nm beneath the surface due to the low energy of the FIB. This close to the surface fabrication is also limiting the spin coherence time of the qubits \cite{BRERETON2020114014} and ultimately can also affect their spectral photostability. Another approach to fabricate these emitters deeper in the material is based on focused protons micro-beam \cite{kraus2017three, ohshima2018creation}, which is however more suitable for ensemble of spins creation. Generally ions implantation produces residual damage that can be detrimental for the preservation of spin coherence of the spin-qubits.
Direct laser fabrication is another maskless approach to create color centres that can have a better control for deeper color centres fabrication. Laser writing has been used to create the $\rm V_{Si}^-$ in 4H-SiC at the single defect level \cite{chen2019laser} using femtosecond laser single pulses and by using a nanosecond laser writing and annealing, the $\rm V_{Si}^-$ has been created directly in photonic crystal cavity\cite{day2023laser}. Ensemble of $\rm V_{Si}^-$ and $\rm V_{Si}\rm V_{C}^0$ were also demonstrated using femtosecond laser writing \cite{castelletto2018photoluminescence, castelletto2021color, almutairi2022direct}. It has been found that the thermal annealing for $\rm V_{Si}\rm V_{C}^0$ in femtosecond laser written samples, yields higher concentration of colour centers at lower annealing temperature of 800$^{\circ}$ C compared to electrons and ions irradiated samples. Due to the low PL emission of the laser fabricated $\rm V_{Si}^-$, attributed to quenching effects due to carbon vacancies, only the ODMR was demonstrated so far for the laser fabricated color centres in SiC. 
Therefore it is currently unknown if the laser fabrication affects these color centres spin coherence properties.

In this paper we directly fabricate large ensemble of telecom O-band photo-emission tentatively associated to the $\rm N_{C} V_{Si}^-$ centre and small ensemble of few divancancies in 4H-SiC. We also show the spin manipulation of the laser written ensemble divacancies, providing relevant information about the preservation of their spin coherence after laser fabrication.

\begin{figure}[b]
\centering
\includegraphics{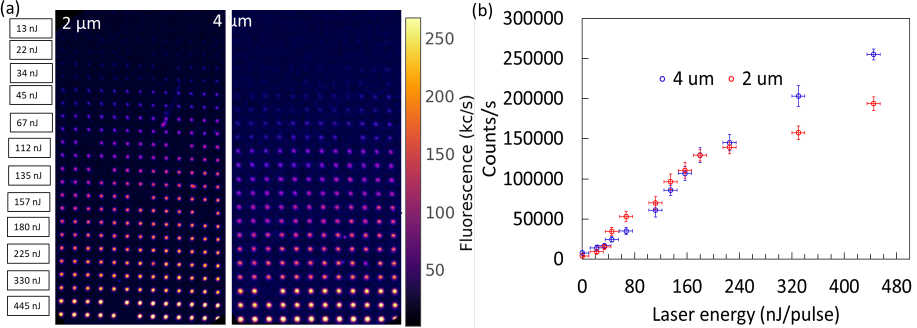}
\caption{(a) RT confocal map of after-annealing laser written array using a 976 nm excitation and 210 $\mu$W optical power. The laser energy per pulse inscription is superimposed. (b) Average photon count rate for 210 $\mu$W optical power after long pass filter at 1000 nm  versus the laser writing energy. The lowest energy detected is corresponding to dots written at 22 nJ/pulse due to high background counts in the non-irradiated area of 5500 cts/s.}
\label{confocal after annealing}
\end{figure}

\section{Material and Methods} 
\subsection{Laser Inscription}

A solid state Yb:KGW laser (Pharos, Light Conversion, Ltd., Vilnius, Lithuania) at 515 nm, 230 fs duration, and repetition rate of 200 kHz was used as described in ref.\cite{castelletto2021color}, as previously utilised to create ensemble of $\rm V_{Si}^-$.  The laser writing was applied to commercial highly N-doped (1$\times$10$^{19}$ cm$^{-3}$) 4H-SiC substrate purchased from SiCrysta and large area array with dots were written with energy per pulse in each line from 2 nJ/pulse to 445 nJ/pulse at 2 and 4 $\mu$m depths. The dots were written 5 $\mu$m apart. 
The writing objective lens has a 100 $\times$ magnification and numerical aperture NA= 0.90 with a working distance of 1.3 mm.
Before laser writing, the SiC sample was cleaned with a Piranha solution to remove organic contaminants, and then annealed in an Argon atmosphere at 1000 °C for 1 h to reduce the as-grown $\rm V_{Si}^-$. 
After laser writing the sample was cleaned with a Piranha solution, and then imaged and the written area spectroscopy performed at room temperature (RT). Subsequently, the sample was annealed at 900 $^\circ$C for 30 minutes in Argon atmosphere. RT and low temperature imaging and spectroscopy were performed on the annealed sample.
\subsection{Confocal imaging, spectroscopy and lifetime measurements}
For imaging the high energy laser written locations (450 nJ to 22 nJ) a home-built scanning fluorescence confocal system equipped with InGaAs single photon counting modules from IDQuantiue (IDQ230) with an approximately 15 \% quantum efficiency and $\approx 2000$ counts/s dark counts was used. The excitation laser was a CW 976 nm diode, and the fluorescence was detected after a dichroic mirror at 980 nm (Di02-R980-25x36) and a long bandpass filter at 1000 nm (Thorlabs, FELH1000). The confocal can operate at RT and at cryogenic temperature as it is equipped with a Montana cryo-station, that can reach 5 K. The confocal was also connected with a Princeton Spectrometer equipped with a PyLoN-IR camera operating at liquid nitrogen temperature used to measure the PL from the laser written array. For room temperature operation an  Olympus dry objective LCPLN-IR (100$\times$, 0.85 NA) was used and for cryogenic operation an dry IR Olympus objective
(60$\times$, 0.65 NA) was used with longer focal distance. 
The objectives were mounted on a PI XYZ computer-controlled stage with an XYZ closed-loop positioner with a 200 $\mu$m travel distance in each direction and step size resolution of 1 nm. The samples were mounted on an XYZ manual stage for room temperature imaging or on an attocube XYZ stage inside the cryostat.
Lifetime measurements were performed with a variable wavelength supercontinuum NKT Photonics Fianium WhiteLase pulsed laser at 800 nm and 40 MHz repetition rate. A time correlator card (PicoQuant GmbH, Berlin Germany TimeHarp 260) was used to obtain time-resolved PL decay traces.
\begin{figure}[tb]
\centering
\includegraphics{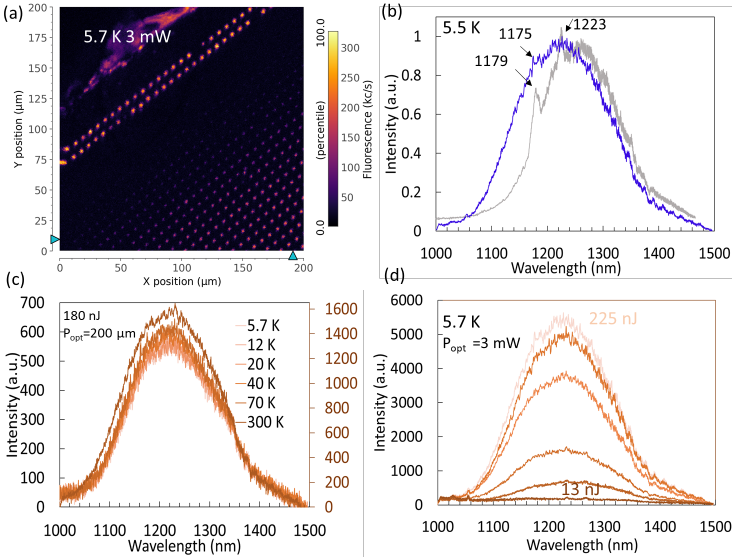}
\caption{(a) Low temperature confocal map of after-annealing laser written array using a 976 nm excitation and 3 mW optical power. (b) Comparison of the 5.5 K PL of a laser written dot at 225 nJ/pulse excited at 3 mW (blue line) compared to a 5 mW 976 nm excitation of a H$^{+}$ irradiated sample at fluence of 10$^{16}$ cm$^{-2}$, fabricated in the same material as per ref. \cite{CastellettoDeterministic2019} after 900 $^\circ$ C annealing (gray line). The ensemble of the $\rm N_{C} V_{Si}^{-}$ was created, where some of the ZPLs are shown. (c) Temperature dependent PL of a 180 nJ/pulse laser written dot excited with 210 $\mu$W optical power, showing a moderate temperature dependence and a PL reduction at lower temperature. (d) Low temperature PL of various laser written dots with energy 225, 180, 157, 67, 34, 13 nJ/pulse excited with 976 nm 3 mW optical power.}
\label{cryo confocal and spectra}
\end{figure}

\subsection{Spin manipulation}

To probe the ODMR and spin coherence of divacancy in correspondence to the low energy laser inscriptions (2-13 nJ) a home-built scanning confocal microscope with an objective with an NA of 0.85 (Olympus, LCPLN-R, 100X) was used. In all of the optical measurements, a 914-nm CW laser, filtered by a shortpass filter (Thorlabs, FESH950), was used to excite those colour centres. A dichroic beamsplitter (Semrock, Di02-R980-25×36) was then used to separate the laser and fluorescence signals. For various measurements at room temperature, the SiC sample was mounted on a closed cycle three-axis piezoelectric stage (PI, E-727.3SD). The fluorescence signals filtered by a 1000-nm longpass filter (Thorlabs, FELH1000) were coupled to a single-mode fibre and then guided to a superconducting nanowire single-photon detector (SNSPD, PHOTEC) with an approximately 85\% quantum efficiency. The number of photons was recorded by a counter (NI, USB-6341).   
For the ODMR, Rabi, Ramsey and spin echo measurements, the microwave sequences were generated using a synthesized signal generator (Mini-Circuits, SSG-6000 RC) and then gated by a switch (Mini-Circuits, ZASWA-2-50DR+). After amplification (Mini-Circuits, ZHL-25W-272+), the microwave signals were fed to a 20-$\mu$m-wide copper wire above the surface of the 4H-SiC sample. The exciting 914-nm CW laser was modulated using an acousto-optic modulator. The timing sequence of the electrical signals for manipulating and synchronizing the laser, microwave and counter was generated using a pulse generator (SpinCore, PBESR-PRO500).
\section{Results and discussion}
\subsection{High energy laser fabrication photo-luminescence}
The 4H-SiC laser written array was inspected before and after annealing. The fluorescent maps before annealing show low intensity PL only from the written array at 2 $\mu m$ from the surface, while no PL was detected from the 4 $\mu m$ depth. 
The RT PL spectra before annealing show that some emitters are present in the sample attributed to surface defects as the intensity is low and comparable with the Raman line and not observed deeper in the sample. Details of the fluorescence map and spectra before annealing are discussed in the Supplementary$^\dag$ Fig. S1.
After annealing the array of dots emission shifts towards a broad brighter emission centred at 1223 nm, present at both laser inscription depths.

\begin{figure*}[tb]
\centering
\includegraphics{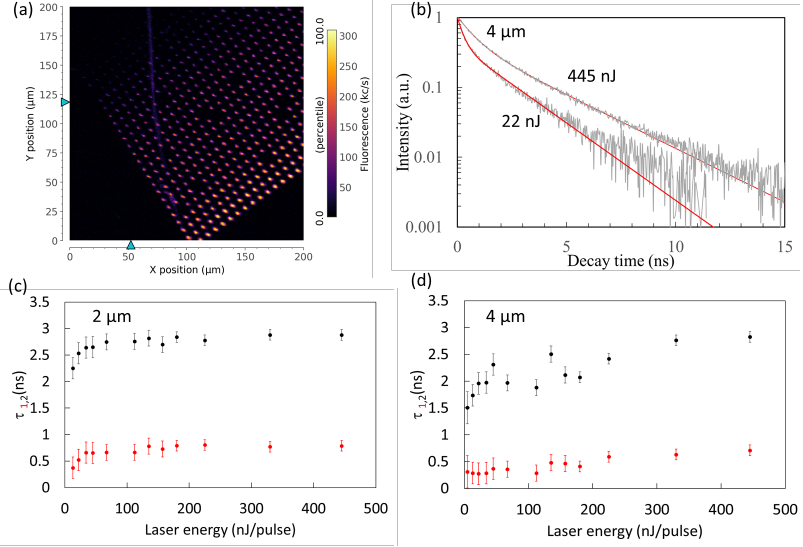}
\caption{(a) RT confocal map of laser written array at 2 $\mu$m using a 800 nm pulsed excitation and 0.5 mW optical power. (b) Exemplary PL time trace for a 445 nJ/pulse and 22 nJ/pulse laser written dots at 4 $\mu$m. (c-d) Lifetime dependence ($\tau_1$ and $\tau_2$ shorter and longer components) versus laser fabrication energy at the two depths.}
\label{lifetime}
\end{figure*}

\begin{figure*}[tb]
\centering
\includegraphics[width=12cm]{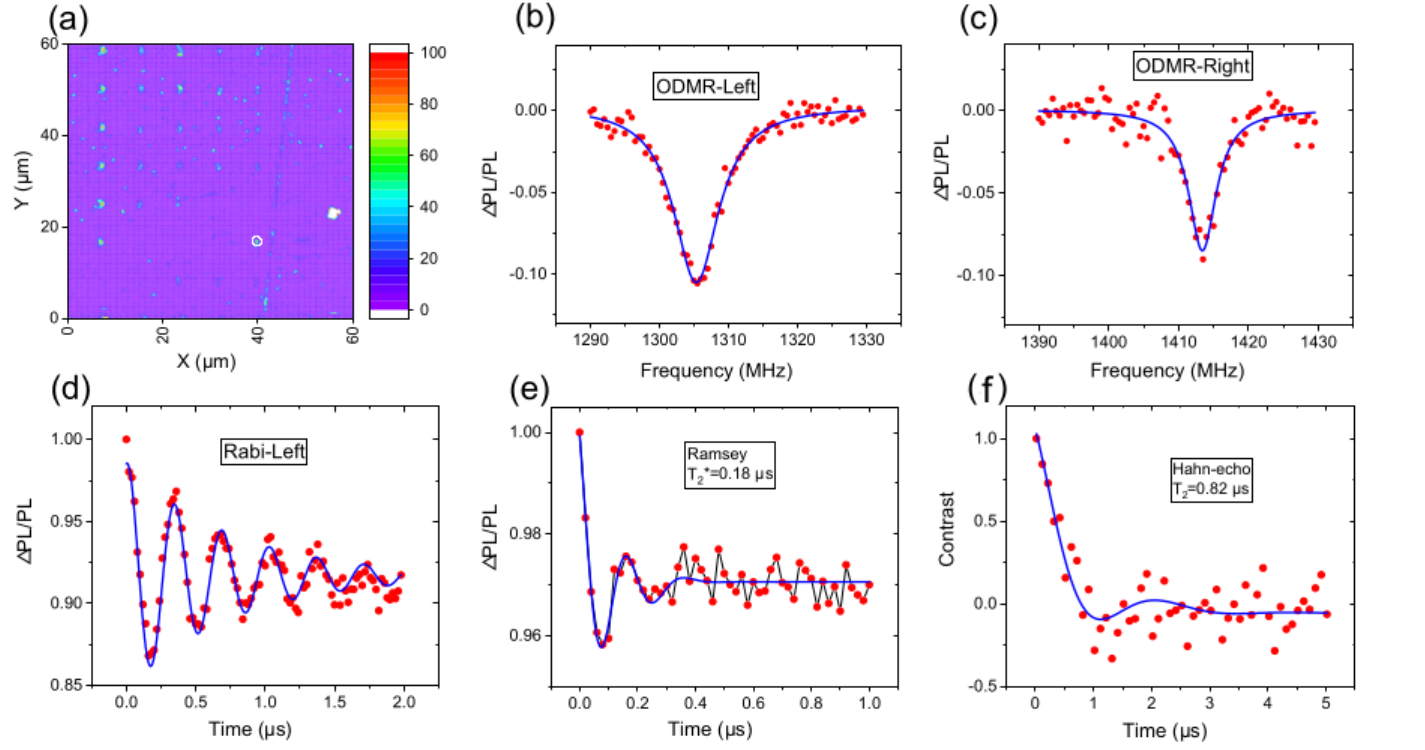}
\caption{(a) $60\times 60~\mu$m$^2$ RT confocal map of laser written array at 4 $\mu$m using a 914 nm CW excitation at 0.2 mW and SNSPD to measure lower energy irradiation from 2 nJ/pulse to 22 nJ/pulse. The circle at 4.5 nJ/pulse indicates a PL5 divacancy identified by the subsequent spin control. (b-c) RT ODMR of the PL5 showing high contrast of 12\% and two branches (Right and left). (d-e-f) Rabi oscillation, Ramsey decay and Hahn-Echo of the PL5's left ODMR branch with related T$_2^{*}$ and T$_2$ measurements.}
\label{PL5 spin control 1}
\end{figure*}
\begin{figure*}[tb]
\centering
\includegraphics[width=12cm]{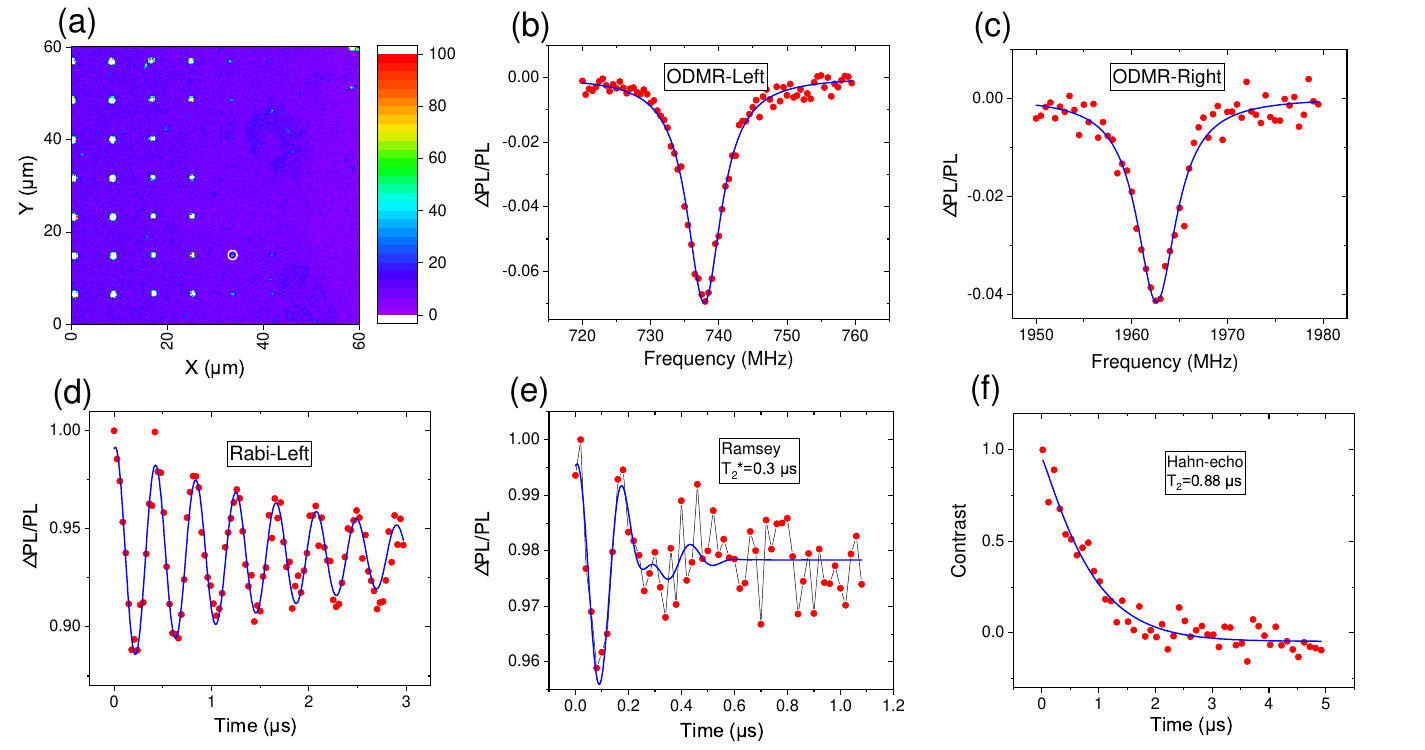}
\caption{(a) $60\times 60~\mu$m$^2$ RT confocal map of laser written array at 2 $\mu$m using a 914 nm CW excitation at 0.2 mW and SNSPD to measure lower energy irradiation from 2 nJ/pulse to 22 nJ/pulse. The circle at 4.5 nJ/pulse indicates a PL6 divacancy identified by the subsequent spin control. (b-c) RT ODMR of the PL6 showing high contrast of 7\% and two branches (Right and left). (d,e,f) Rabi oscillation, Ramsey decay and Hahn-Echo of the PL6's left ODMR branch with related T$_2^{*}$ and T$_2$ measurements.}
\label{PL6 spin control}
\end{figure*}
In Fig.\ref{confocal after annealing} (a) RT confocal images of the laser written array at two depths are shown with a strong PL, which is increasing with laser energy irradiation, as shown in Fig.\ref{confocal after annealing} (b). The closer to the surface laser fabrication shows a reduction of PL for dots with higher fabrication energy.

Low temperature confocal scan and PL spectra are shown in Fig.\ref{cryo confocal and spectra}. The laser written dots from 445 nJ to 13 nJ fabrication energy all exhibit a bright PL centred around 1223 nm at 5.7 K even with 210 $\mu$W optical excitation power, and by exciting at high optical power of 3.3 mW a small peak at 1175 nm is emerging Fig.\ref{cryo confocal and spectra}(b). The low temperature PL of a laser written dot is compared to PL measurement with the same cryogenic station of a proton irradiated sample of the same N-doped 4H-SiC, as described in ref.  \cite{CastellettoDeterministic2019}. Here the ZPLs at 1179 nm and 1223 nm are corresponding to the $\rm N_{C} V_{Si}^-$ ensemble. The laser irradiated dots low temperature PL appears blue shifted and the ZPLs are less evident. The broad 1223 nm centred emission created by laser writing could be associated to the fabrication of the $\rm N_{C} V_{Si}^-$ in ensemble, however the lack of a strong temperature dependence of the emission, as shown in Fig.\ref{cryo confocal and spectra} (c), suggests the creation of additional damage-related defects quenching the $\rm N_{C} V_{Si}^-$ ZPLs. Fig.\ref{cryo confocal and spectra} (d) showcases the low-temperature PL characteristics of various laser-written dots across differing energy levels. 

A temperature annealing study should be performed to optimise the formation of the $\rm N_{C} V_{Si}^-$ ZPLs in the laser written sample and reduce the additional damage related emission, quenching the color centres. It is to be noted that the same laser fabrication studies post annealing were performed in HPSI Research Grade Semi-insulating SiC from CREE (N doping $<$5$\times$10$^{14}$ cm$^{-3}$) and no PL centres at 1223 nm were observed, suggesting that this emission is N-doping related, as such further studies in different nitrogen doping samples are needed. Finally lower energy fabrication dots should be further investigated using more sensitive detectors in the spectral region of the $\rm N_{C} V_{Si}^-$.
To further investigate the origin the the laser fabricated emission centre at 1223 nm, lifetime measurements were performed using a pulsed 800 nm laser excitation and the lifetime measured for different laser inscription energy from the dots in the confocal map in Fig.\ref{lifetime} (a). The fluorescence time traces are fitted by two exponential decays, as shown in Fig.\ref{lifetime} (b) for two exemplary dots and the short and long optical lifetime are summarized in Fig.\ref{lifetime} (c-d) for the two laser irradiation depths, showing lower lifetime for the 4$\mu$m depth from 1.5$\pm$0.3 to 2.8$\pm$0.1 ns compared to the 2 $\mu$m from 2.3$\pm$ 0.2 to 2.9$\pm$0.1 ns. It has been reported that the optical lifetime of single $\rm N_{C} V_{Si}^-$ in 4H-SiC is in similar ranges, specifically 2.7 $\pm$ 0.2 and 2.3 $\pm$ 0.2 ns \cite{Wangexperimental2020}.
\subsection{Low energy laser fabrication spin control}
Long spin coherence of spin-photon interfaces in solid is relevant to avoid loss of phase information in quantum bits and to improve sensitivity in quantum sensing. Spin relaxation times ($T_x$)  can be measured using a spin manipulation mechanism \cite{wolfowicz2021quantum}. Solid-state spin qubits phase coherence is limited by surrounding fluctuating nuclear and electron spins, and generally fabrication in the material included their formation can induce spin-decoherence. As such it is important to test the fabricated color centres spin properties. The spin decoherence is distinguished in two main components: the inhomogeneous dephasing time $T^*_2$, which is intrinsic to the spin qubit,  measured by Ramsey interferometry and the homogeneous dephasing time $T_2$ measured by Hahn-Echo sequence.  
 
 By inspecting the lower energy fabrication dots at 2-13 nJ/pulse with superconducting nanowire single photon detectors, the ODMR and spin manipulation of several divacancies were observed. PL5 (1043 nm) was identified in 2 and 4 $\mu$m laser irradiated dots with  4.5 nJ/pulse fabrication, as shown from the confocal map in Fig. \ref{PL5 spin control 1}(a) for the 4  $\mu$m dot. The ODMR spectra in Figs. \ref{PL5 spin control 1} (b) and (c) show two branches in absence of magnetic field at 1305.3 MHz (left) and 1413.4 MHz (right), corresponding to about 38 MHz frequency shift compared to the 1343.7 MHz (left) and 1375.7 MHz (right) observed in ref. \cite{li2022room}. 
 The frequency shift might be attributed to the strain effect. The spectral broadening of the two branches ODMR obtained from a Lorentzian-shaped multi peaks fitting are 7 and 5 MHz. 
 Rabi oscillation, Ramsey decay and Hanh-Echo spin control were performed on the PL5 left branch as shown in Fig.\ref{PL5 spin control 1}(d,e,f) providing a spin dephasing time of 180 ns for Ramsey and 820 ns for Hanh-Echo. The measured T$_2$=820 ns is lower than the PL5 spin coherence of 24$\pm$2.4 $\mu$s, achieved by carbon ion (C$^{+}$) implantation and annealing in a 12.5-$\mu$m-thick epitaxial layer of single-crystal 4H-SiC with a nitrogen doping density of 5 $\times$ 10$^{15}$ cm$^{-3}$ \cite{li2022room}. The lower coherence time measured here for the PL5 compared to this previous study can be primarily due to the high nitrogen concentration of the sample compared to previous fabrications. In addition it is not excluded that $\rm N_{C} V_{Si}^-$ could be created also at this lower laser energy fabrication, which provides background for the divacancy spin read-out in this study. Longer wavelength excitation at 1064 nm and tuning of the annealing temperature should be used to selectively activate only the $\rm N_{C} V_{Si}^-$. 

PL6 divacancies (1038 nm) were also observed in the 2 and 4 $\mu$m and 4.5 and 13 nJ/pulse laser fabrication as shown for the 4.5 nJ PL6 in Fig. \ref{PL6 spin control} (a). The ODMR of the left and right branches are displayed in Fig. \ref{PL6 spin control} (b,c) in the presence of an external magnetic field of 219 G, showing a signal to noise contrast of  7\% ; the Rabi oscillation, Ramsey decay and Hahn-Echo spin control are displayed in Fig. \ref{PL6 spin control} (d-f), showing slightly longer spin coherence properties of the PL5, lower however than few microseconds as demonstrated in ion implanted low nitrogen SiC.

 
 As the divacancies spin manipulation was achieved in correspondence to 4.5 nJ and 13 nJ, these results suggest that the laser fabrication in this case is in the multiphoton absorption (MPI) regime and able to achieve close to few defects scale manufacturing; here the energy per pulse is well below the minimum threshold energy pulse (minimum fluence) to create vacancies, as previously estimated\cite{chen2019laser}, which is for this laser (515 nm, $NA=0.9$, pulse=230 fs) around 31 nJ/pulse (7~J/cm$^2$). This result is in agreement with the required much lower fluences $<1$~J/cm$^2$ to achieve close to single defect creation and avoid ablation, as estimated using molecular dynamic simulations and two-temperature model~\cite{10119524}.

\section*{Conclusions}
In summary, a bright and photostable telecom O-band emission in 4H-SiC has been created in 4H-SiC with high energy laser irradiation (22 nJ-445 nJ) and subsequent thermal annealing, the emission could be linked to a large ensemble of the $\rm N_{C} V_{Si}^-$, which was not present before laser fabrication. This is inferred from the optical lifetime, while the ZPLs at low temperature are not well discerned. This color centre has been created by laser writing in the N-type 4H-SiC and more studies on doping concentration and thermal annealing in 4H-SiC for the creation of $\rm N_{C} V_{Si}^-$ by direct laser writing are necessary. The high energy laser fabrication provides consistent emission across all energy fluence. By tuning the laser fabrication energy and annealing, a better control of this single color centre formation can be achieved. The laser writing at low energy and post thermal annealing in the same sample also created few divacancies of the type PL5, PL6, which were not present before laser fabrication, neither in the pristine areas. These divacancies were identified by their ODMR signature and for the first time the coherent spin control was applied to laser written divacancies in 4H-SiC. While the divacancies formation yield can be increased by better tuning the thermal annealing temperature and duration, the spin coherence time can be increased by laser writing the divacancies in lower nitrogen concentration SiC samples, as previously demonstrated using ions implantation. This work stems for the first time spin control of laser fabricated SiC spin defects and conveys that the spin coherence of divacancies is preserved after laser fabrication when using low energy fabrication well below the ablation threshold. Therefore direct laser writing is promising if finely tuned at low energy irradiation to achieve better spin coherence of divacancy color centres in SiC and localise them for inscription of spin qubits in photonics structure. By laser writing at low energy it will be possible to achieve for spin-photon interface with subdiffraction spatial resolution localisation\cite{wang2024laser}. Conversely higher energy laser fabrication can be used for creating an ensemble of localised telecon O-band emitters for biomedical fluorescent labeling and quantum sensing \cite{kucsko2013nanometre}.
\section{Supporting Information}
 The supporting information file shows the laser inscription confocal map and photoluminescence before thermal annealing.
\section{Author contributions}
S.C.,J.-S.X: Conceptualization, Methodology. J.M.,T.K.,J.S.:Fabrication. Z.-H.H.,Z.-X.H.,S.C.,J.-S.X.: Data acquisition, curation, Writing- Original draft preparation. Z.-H.H,Z.-X.H.,S.C.,J.-S.X: Visualization, Investigation. J.-S.X.,C.-F.L.,G.-C.G.,J.S.: Supervision. S.C.,J.-S.X.: Writing- Reviewing and Editing.
\section{Conflicts of interest}
There are no conflicts to declare.
\section{Data availability}
Data are available from the corresponding authors upon request.
\section{Acknowledgments}
J.-S.X acknowledges support from the Innovation Program for Quantum Science and Technology (2021ZD0301400), and the National Natural Science Foundation of China (Grants Nos. 92365205 and 11821404).
S.J. acknowledges partial support by the ARC Discovery DP240103231 grant.
S.J. and T.K. are grateful to Workshop of Photonics (WoP), Vilnius, Lithuania, for the technology transfer project used to establish industrial fs-laser machining at Swinburne. S.C. acknowledges Phillip Reineick for providing laboratory assistance for sample annealing. S.C. acknowledges the financial laboratory support from the ARC Centre of Excellence for Nanoscale BioPhotonics (No. CE140100003), and the LIEF scheme grant (no. LE140100131). This work was partially performed at the University of Science and Technology of China Center for Micro and Nanoscale Research and Fabrication.
\bibliography{SiCreferences}

\providecommand{\latin}[1]{#1}
\makeatletter
\providecommand{\doi}
  {\begingroup\let\do\@makeother\dospecials
  \catcode`\{=1 \catcode`\}=2 \doi@aux}
\providecommand{\doi@aux}[1]{\endgroup\texttt{#1}}
\makeatother
\providecommand*\mcitethebibliography{\thebibliography}
\csname @ifundefined\endcsname{endmcitethebibliography}  {\let\endmcitethebibliography\endthebibliography}{}
\begin{mcitethebibliography}{43}
\providecommand*\natexlab[1]{#1}
\providecommand*\mciteSetBstSublistMode[1]{}
\providecommand*\mciteSetBstMaxWidthForm[2]{}
\providecommand*\mciteBstWouldAddEndPuncttrue
  {\def\EndOfBibitem{\unskip.}}
\providecommand*\mciteBstWouldAddEndPunctfalse
  {\let\EndOfBibitem\relax}
\providecommand*\mciteSetBstMidEndSepPunct[3]{}
\providecommand*\mciteSetBstSublistLabelBeginEnd[3]{}
\providecommand*\EndOfBibitem{}
\mciteSetBstSublistMode{f}
\mciteSetBstMaxWidthForm{subitem}{(\alph{mcitesubitemcount})}
\mciteSetBstSublistLabelBeginEnd
  {\mcitemaxwidthsubitemform\space}
  {\relax}
  {\relax}

\bibitem[Casady and Johnson(1996)Casady, and Johnson]{casady1996status}
Casady,~J.; Johnson,~R.~W. Status of silicon carbide (SiC) as a wide-bandgap semiconductor for high-temperature applications: A review. \emph{Solid-State Electronics} \textbf{1996}, \emph{39}, 1409--1422\relax
\mciteBstWouldAddEndPuncttrue
\mciteSetBstMidEndSepPunct{\mcitedefaultmidpunct}
{\mcitedefaultendpunct}{\mcitedefaultseppunct}\relax
\EndOfBibitem
\bibitem[Lukin \latin{et~al.}(2020)Lukin, Guidry, and Vu{\v{c}}kovi{\'c}]{lukin2020integrated}
Lukin,~D.~M.; Guidry,~M.~A.; Vu{\v{c}}kovi{\'c},~J. Integrated quantum photonics with silicon carbide: challenges and prospects. \emph{PRX Quantum} \textbf{2020}, \emph{1}, 020102\relax
\mciteBstWouldAddEndPuncttrue
\mciteSetBstMidEndSepPunct{\mcitedefaultmidpunct}
{\mcitedefaultendpunct}{\mcitedefaultseppunct}\relax
\EndOfBibitem
\bibitem[Castelletto \latin{et~al.}(2022)Castelletto, Peruzzo, Bonato, Johnson, Radulaski, Ou, Kaiser, and Wrachtrup]{castelletto2022silicon}
Castelletto,~S.; Peruzzo,~A.; Bonato,~C.; Johnson,~B.~C.; Radulaski,~M.; Ou,~H.; Kaiser,~F.; Wrachtrup,~J. Silicon Carbide Photonics Bridging Quantum Technology. \emph{ACS Photonics} \textbf{2022}, \emph{9}, 1434--1457\relax
\mciteBstWouldAddEndPuncttrue
\mciteSetBstMidEndSepPunct{\mcitedefaultmidpunct}
{\mcitedefaultendpunct}{\mcitedefaultseppunct}\relax
\EndOfBibitem
\bibitem[Bader \latin{et~al.}(2024)Bader, Arianfard, Peruzzo, and Castelletto]{bader2024analysis}
Bader,~J.; Arianfard,~H.; Peruzzo,~A.; Castelletto,~S. Analysis, recent challenges and capabilities of spin-photon interfaces in Silicon carbide-on-insulator. \emph{npj Nanophotonics} \textbf{2024}, \emph{1}, 29\relax
\mciteBstWouldAddEndPuncttrue
\mciteSetBstMidEndSepPunct{\mcitedefaultmidpunct}
{\mcitedefaultendpunct}{\mcitedefaultseppunct}\relax
\EndOfBibitem
\bibitem[Wolfowicz \latin{et~al.}(2021)Wolfowicz, Heremans, Anderson, Kanai, Seo, Gali, Galli, and Awschalom]{wolfowicz2021quantum}
Wolfowicz,~G.; Heremans,~F.~J.; Anderson,~C.~P.; Kanai,~S.; Seo,~H.; Gali,~A.; Galli,~G.; Awschalom,~D.~D. Quantum guidelines for solid-state spin defects. \emph{Nat. Rev. Mater.} \textbf{2021}, \emph{6}, 906–925\relax
\mciteBstWouldAddEndPuncttrue
\mciteSetBstMidEndSepPunct{\mcitedefaultmidpunct}
{\mcitedefaultendpunct}{\mcitedefaultseppunct}\relax
\EndOfBibitem
\bibitem[Anderson \latin{et~al.}(2022)Anderson, Glen, Zeledon, Bourassa, Jin, Zhu, Vorwerk, Crook, Abe, Ul-Hassan, Ohshima, Son, Galli, and Awschalom]{anderson2021five}
Anderson,~C.~P.; Glen,~E.~O.; Zeledon,~C.; Bourassa,~A.; Jin,~Y.; Zhu,~Y.; Vorwerk,~C.; Crook,~A.~L.; Abe,~H.; Ul-Hassan,~J.; Ohshima,~T.; Son,~N.~T.; Galli,~G.; Awschalom,~D.~D. Five-second coherence of a single spin with single-shot readout in silicon carbide. \emph{Sci. Adv.} \textbf{2022}, \emph{8}, eabm5912\relax
\mciteBstWouldAddEndPuncttrue
\mciteSetBstMidEndSepPunct{\mcitedefaultmidpunct}
{\mcitedefaultendpunct}{\mcitedefaultseppunct}\relax
\EndOfBibitem
\bibitem[Nagy \latin{et~al.}(2019)Nagy, Niethammer, Widmann, Chen, Udvarhelyi, Bonato, Hassan, Karhu, Ivanov, Son, Maze, Ohshima, Soykal, Gali, Lee, Kaiser, and Wrachtrup]{Nagy2019}
Nagy,~R. \latin{et~al.}  {High-fidelity spin and optical control of single silicon-vacancy centres in silicon carbide}. \emph{Nat. Commun.} \textbf{2019}, \emph{10}, 1954\relax
\mciteBstWouldAddEndPuncttrue
\mciteSetBstMidEndSepPunct{\mcitedefaultmidpunct}
{\mcitedefaultendpunct}{\mcitedefaultseppunct}\relax
\EndOfBibitem
\bibitem[Son \latin{et~al.}(2020)Son, Anderson, Bourassa, Miao, Babin, Widmann, Niethammer, Ul~Hassan, Morioka, Ivanov, Kaiser, Wrachtrup, and Awschalom]{Son2020perspective}
Son,~N.~T.; Anderson,~C.~P.; Bourassa,~A.; Miao,~K.~C.; Babin,~C.; Widmann,~M.; Niethammer,~M.; Ul~Hassan,~J.; Morioka,~N.; Ivanov,~I.~G.; Kaiser,~F.; Wrachtrup,~J.; Awschalom,~D.~D. Developing silicon carbide for quantum spintronics. \emph{Appl. Phys. Lett.} \textbf{2020}, \emph{116}, 190501\relax
\mciteBstWouldAddEndPuncttrue
\mciteSetBstMidEndSepPunct{\mcitedefaultmidpunct}
{\mcitedefaultendpunct}{\mcitedefaultseppunct}\relax
\EndOfBibitem
\bibitem[Morioka \latin{et~al.}(2020)Morioka, Babin, Nagy, Gediz, Hesselmeier, Liu, Joliffe, Niethammer, Dasari, Vorobyov, \latin{et~al.} others]{morioka2020spin}
Morioka,~N.; Babin,~C.; Nagy,~R.; Gediz,~I.; Hesselmeier,~E.; Liu,~D.; Joliffe,~M.; Niethammer,~M.; Dasari,~D.; Vorobyov,~V.; others Spin-controlled generation of indistinguishable and distinguishable photons from silicon vacancy centres in silicon carbide. \emph{Nat. Commun.} \textbf{2020}, \emph{11}, 1--8\relax
\mciteBstWouldAddEndPuncttrue
\mciteSetBstMidEndSepPunct{\mcitedefaultmidpunct}
{\mcitedefaultendpunct}{\mcitedefaultseppunct}\relax
\EndOfBibitem
\bibitem[Fang \latin{et~al.}(2024)Fang, Lai, Li, Su, Lu, Yang, Liu, Qiao, Li, He, Huang, Li, You, Huo, Bao, and Pan]{Fang2024spin_photon}
Fang,~R.-Z. \latin{et~al.}  Experimental Generation of Spin-Photon Entanglement in Silicon Carbide. \emph{Phys. Rev. Lett.} \textbf{2024}, \emph{132}, 160801\relax
\mciteBstWouldAddEndPuncttrue
\mciteSetBstMidEndSepPunct{\mcitedefaultmidpunct}
{\mcitedefaultendpunct}{\mcitedefaultseppunct}\relax
\EndOfBibitem
\bibitem[Bourassa \latin{et~al.}(2020)Bourassa, Anderson, Miao, Onizhuk, Ma, Crook, Abe, Ul-Hassan, Ohshima, Son, \latin{et~al.} others]{bourassa2020entanglement}
Bourassa,~A.; Anderson,~C.~P.; Miao,~K.~C.; Onizhuk,~M.; Ma,~H.; Crook,~A.~L.; Abe,~H.; Ul-Hassan,~J.; Ohshima,~T.; Son,~N.~T.; others Entanglement and control of single nuclear spins in isotopically engineered silicon carbide. \emph{Nat. Mater.} \textbf{2020}, \emph{19}, 1319--1325\relax
\mciteBstWouldAddEndPuncttrue
\mciteSetBstMidEndSepPunct{\mcitedefaultmidpunct}
{\mcitedefaultendpunct}{\mcitedefaultseppunct}\relax
\EndOfBibitem
\bibitem[Koehl \latin{et~al.}(2011)Koehl, Buckley, Heremans, Calusine, and Awschalom]{koehl_room_2011}
Koehl,~W.~F.; Buckley,~B.~B.; Heremans,~F.~J.; Calusine,~G.; Awschalom,~D.~D. Room temperature coherent control of defect spin qubits in silicon carbide. \emph{Nature} \textbf{2011}, \emph{479}, 84--87\relax
\mciteBstWouldAddEndPuncttrue
\mciteSetBstMidEndSepPunct{\mcitedefaultmidpunct}
{\mcitedefaultendpunct}{\mcitedefaultseppunct}\relax
\EndOfBibitem
\bibitem[Falk \latin{et~al.}(2013)Falk, Buckley, Calusine, Koehl, Dobrovitski, Politi, Zorman, Feng, and Awschalom]{falk_polytype_2013}
Falk,~A.~L.; Buckley,~B.~B.; Calusine,~G.; Koehl,~W.~F.; Dobrovitski,~V.~V.; Politi,~A.; Zorman,~C.~A.; Feng,~P. X.-L.; Awschalom,~D.~D. Polytype control of spin qubits in silicon carbide. \emph{Nat. Commun.} \textbf{2013}, \emph{4}, 1819\relax
\mciteBstWouldAddEndPuncttrue
\mciteSetBstMidEndSepPunct{\mcitedefaultmidpunct}
{\mcitedefaultendpunct}{\mcitedefaultseppunct}\relax
\EndOfBibitem
\bibitem[Christle \latin{et~al.}(2015)Christle, Falk, Andrich, Klimov, Hassan, Son, Janzén, Ohshima, and Awschalom]{christle_isolated_2015}
Christle,~D.~J.; Falk,~A.~L.; Andrich,~P.; Klimov,~P.~V.; Hassan,~J.~U.; Son,~N.~T.; Janzén,~E.; Ohshima,~T.; Awschalom,~D.~D. Isolated electron spins in silicon carbide with millisecond coherence times. \emph{Nat. Mater.} \textbf{2015}, \emph{14}, 160--163\relax
\mciteBstWouldAddEndPuncttrue
\mciteSetBstMidEndSepPunct{\mcitedefaultmidpunct}
{\mcitedefaultendpunct}{\mcitedefaultseppunct}\relax
\EndOfBibitem
\bibitem[Klimov \latin{et~al.}(2015)Klimov, Falk, Christle, Dobrovitski, and Awschalom]{klimov_quantum_2015}
Klimov,~P.~V.; Falk,~A.~L.; Christle,~D.~J.; Dobrovitski,~V.~V.; Awschalom,~D.~D. Quantum entanglement at ambient conditions in a macroscopic solid-state spin ensemble. \emph{Sci. Adv.} \textbf{2015}, \emph{1}, e1501015\relax
\mciteBstWouldAddEndPuncttrue
\mciteSetBstMidEndSepPunct{\mcitedefaultmidpunct}
{\mcitedefaultendpunct}{\mcitedefaultseppunct}\relax
\EndOfBibitem
\bibitem[Zwier \latin{et~al.}(2015)Zwier, O’Shea, Onur, and van~der Wal]{zwier_all-optical_2015}
Zwier,~O.~V.; O’Shea,~D.; Onur,~A.~R.; van~der Wal,~C.~H. All-optical coherent population trapping with defect spin ensembles in silicon carbide. \emph{Sci. Rep.} \textbf{2015}, \emph{5}, 10931\relax
\mciteBstWouldAddEndPuncttrue
\mciteSetBstMidEndSepPunct{\mcitedefaultmidpunct}
{\mcitedefaultendpunct}{\mcitedefaultseppunct}\relax
\EndOfBibitem
\bibitem[Zhou \latin{et~al.}(2021)Zhou, Li, Hao, Yan, Yang, Wang, Lin, Liu, Liu, Li, \latin{et~al.} others]{zhou2021experimental}
Zhou,~J.-Y.; Li,~Q.; Hao,~Z.-Y.; Yan,~F.-F.; Yang,~M.; Wang,~J.-F.; Lin,~W.-X.; Liu,~Z.-H.; Liu,~W.; Li,~H.; others Experimental determination of the dipole orientation of single color centers in silicon carbide. \emph{ACS Photonics} \textbf{2021}, \emph{8}, 2384--2391\relax
\mciteBstWouldAddEndPuncttrue
\mciteSetBstMidEndSepPunct{\mcitedefaultmidpunct}
{\mcitedefaultendpunct}{\mcitedefaultseppunct}\relax
\EndOfBibitem
\bibitem[Li \latin{et~al.}(2022)Li, Wang, Yan, Zhou, Wang, Liu, Guo, Zhou, Gali, Liu, \latin{et~al.} others]{li2022room}
Li,~Q.; Wang,~J.-F.; Yan,~F.-F.; Zhou,~J.-Y.; Wang,~H.-F.; Liu,~H.; Guo,~L.-P.; Zhou,~X.; Gali,~A.; Liu,~Z.-H.; others Room-temperature coherent manipulation of single-spin qubits in silicon carbide with a high readout contrast. \emph{National Science Review} \textbf{2022}, \emph{9}, nwab122\relax
\mciteBstWouldAddEndPuncttrue
\mciteSetBstMidEndSepPunct{\mcitedefaultmidpunct}
{\mcitedefaultendpunct}{\mcitedefaultseppunct}\relax
\EndOfBibitem
\bibitem[Castelletto \latin{et~al.}(2023)Castelletto, Lew, Lin, and Xu]{Castelletto_2024}
Castelletto,~S.; Lew,~C. T.-K.; Lin,~W.-X.; Xu,~J.-S. Quantum systems in silicon carbide for sensing applications. \emph{Reports on Progress in Physics} \textbf{2023}, \emph{87}, 014501\relax
\mciteBstWouldAddEndPuncttrue
\mciteSetBstMidEndSepPunct{\mcitedefaultmidpunct}
{\mcitedefaultendpunct}{\mcitedefaultseppunct}\relax
\EndOfBibitem
\bibitem[Widmann \latin{et~al.}(2014)Widmann, Lee, Rendler, Son, Fedder, Paik, Yang, Zhao, Yang, Booker, Denisenko, Jamali, Momenzadeh, Gerhardt, Ohshima, Gali, Janz\'{e}n, and Wrachtrup]{Widmann2014a}
Widmann,~M. \latin{et~al.}  {Coherent control of single spins in silicon carbide at room temperature.} \emph{Nat. Mater.} \textbf{2014}, \emph{14}, 164--168\relax
\mciteBstWouldAddEndPuncttrue
\mciteSetBstMidEndSepPunct{\mcitedefaultmidpunct}
{\mcitedefaultendpunct}{\mcitedefaultseppunct}\relax
\EndOfBibitem
\bibitem[Babin \latin{et~al.}(2022)Babin, St{\"o}hr, Morioka, Linkewitz, Steidl, W{\"o}rnle, Liu, Hesselmeier, Vorobyov, Denisenko, \latin{et~al.} others]{babin2021nanofabricated}
Babin,~C.; St{\"o}hr,~R.; Morioka,~N.; Linkewitz,~T.; Steidl,~T.; W{\"o}rnle,~R.; Liu,~D.; Hesselmeier,~E.; Vorobyov,~V.; Denisenko,~A.; others Fabrication and nanophotonic waveguide integration of silicon carbide colour centres with preserved spin-optical coherence. \emph{Nat. Mater.} \textbf{2022}, \emph{21}, 67--73\relax
\mciteBstWouldAddEndPuncttrue
\mciteSetBstMidEndSepPunct{\mcitedefaultmidpunct}
{\mcitedefaultendpunct}{\mcitedefaultseppunct}\relax
\EndOfBibitem
\bibitem[von Bardeleben \latin{et~al.}(2015)von Bardeleben, Cantin, Rauls, and Gerstmann]{von_bardeleben_identification_2015}
von Bardeleben,~H.~J.; Cantin,~J.~L.; Rauls,~E.; Gerstmann,~U. Identification and magneto-optical properties of the {NV} center in \${4H}{\textbackslash}ensuremath\{-\}{\textbackslash}mathrm\{{SiC}\}\$. \emph{Phys. Rev. B Condens. Matter.} \textbf{2015}, \emph{92}, 064104\relax
\mciteBstWouldAddEndPuncttrue
\mciteSetBstMidEndSepPunct{\mcitedefaultmidpunct}
{\mcitedefaultendpunct}{\mcitedefaultseppunct}\relax
\EndOfBibitem
\bibitem[von Bardeleben \latin{et~al.}(2016)von Bardeleben, Cantin, Csóré, Gali, Rauls, and Gerstmann]{von_bardeleben_nv_2016}
von Bardeleben,~H.~J.; Cantin,~J.~L.; Csóré,~A.; Gali,~A.; Rauls,~E.; Gerstmann,~U. {NV} centers in 3C,4H, and 6H silicon carbide: {A} variable platform for solid-state qubits and nanosensors. \emph{Phys. Rev. B Condens. Matter.} \textbf{2016}, \emph{94}, 121202\relax
\mciteBstWouldAddEndPuncttrue
\mciteSetBstMidEndSepPunct{\mcitedefaultmidpunct}
{\mcitedefaultendpunct}{\mcitedefaultseppunct}\relax
\EndOfBibitem
\bibitem[Zargaleh \latin{et~al.}(2018)Zargaleh, von Bardeleben, Cantin, Gerstmann, Hameau, Eblé, and Gao]{zargaleh_electron_2018}
Zargaleh,~S.~A.; von Bardeleben,~H.~J.; Cantin,~J.~L.; Gerstmann,~U.; Hameau,~S.; Eblé,~B.; Gao,~W. Electron paramagnetic resonance tagged high-resolution excitation spectroscopy of {NV}-centers in 4H-{SiC}. \emph{Phys. Rev. B Condens. Matter.} \textbf{2018}, \emph{98}, 214113\relax
\mciteBstWouldAddEndPuncttrue
\mciteSetBstMidEndSepPunct{\mcitedefaultmidpunct}
{\mcitedefaultendpunct}{\mcitedefaultseppunct}\relax
\EndOfBibitem
\bibitem[Mu \latin{et~al.}(2020)Mu, Zargaleh, von Bardeleben, Fröch, Nonahal, Cai, Yang, Yang, Li, Aharonovich, and Gao]{mu_coherent_2020}
Mu,~Z.; Zargaleh,~S.~A.; von Bardeleben,~H.~J.; Fröch,~J.~E.; Nonahal,~M.; Cai,~H.; Yang,~X.; Yang,~J.; Li,~X.; Aharonovich,~I.; Gao,~W. Coherent Manipulation with Resonant Excitation and Single Emitter Creation of Nitrogen Vacancy Centers in 4H Silicon Carbide. \emph{Nano Lett.} \textbf{2020}, \emph{20}, 6142--6147\relax
\mciteBstWouldAddEndPuncttrue
\mciteSetBstMidEndSepPunct{\mcitedefaultmidpunct}
{\mcitedefaultendpunct}{\mcitedefaultseppunct}\relax
\EndOfBibitem
\bibitem[Wang \latin{et~al.}(2020)Wang, Yan, Li, Liu, Liu, Guo, Guo, Zhou, Cui, Wang, Zhou, Xu, Xu, Li, and Guo]{wang_coherent_2020}
Wang,~J.-F.; Yan,~F.-F.; Li,~Q.; Liu,~Z.-H.; Liu,~H.; Guo,~G.-P.; Guo,~L.-P.; Zhou,~X.; Cui,~J.-M.; Wang,~J.; Zhou,~Z.-Q.; Xu,~X.-Y.; Xu,~J.-S.; Li,~C.-F.; Guo,~G.-C. Coherent {Control} of {Nitrogen}-{Vacancy} {Center} {Spins} in {Silicon} {Carbide} at {Room} {Temperature}. \emph{Phys. Rev. Lett.} \textbf{2020}, \emph{124}, 223601\relax
\mciteBstWouldAddEndPuncttrue
\mciteSetBstMidEndSepPunct{\mcitedefaultmidpunct}
{\mcitedefaultendpunct}{\mcitedefaultseppunct}\relax
\EndOfBibitem
\bibitem[Castelletto \latin{et~al.}(2019)Castelletto, Atem, Inam, von Bardeleben, Hameau, Almutairi, Guillot, ichiro Sato, Boretti, and Bluet]{CastellettoDeterministic2019}
Castelletto,~S.; Atem,~A. S.~A.; Inam,~F.~A.; von Bardeleben,~H.~J.; Hameau,~S.; Almutairi,~A.~F.; Guillot,~G.; ichiro Sato,~S.; Boretti,~A.; Bluet,~J.~M. Deterministic placement of ultra-bright near-infrared color centers in arrays of silicon carbide micropillars. \emph{Beilstein Journal of Nanotechnology} \textbf{2019}, \emph{10}, 2383--2395\relax
\mciteBstWouldAddEndPuncttrue
\mciteSetBstMidEndSepPunct{\mcitedefaultmidpunct}
{\mcitedefaultendpunct}{\mcitedefaultseppunct}\relax
\EndOfBibitem
\bibitem[Wang \latin{et~al.}(2018)Wang, Zhou, Wang, Rasmita, Yang, Li, von Bardeleben, and Gao]{wang_bright_2018}
Wang,~J.; Zhou,~Y.; Wang,~Z.; Rasmita,~A.; Yang,~J.; Li,~X.; von Bardeleben,~H.~J.; Gao,~W. Bright room temperature single photon source at telecom range in cubic silicon carbide. \emph{Nat. Commun.} \textbf{2018}, \emph{9}, 4106\relax
\mciteBstWouldAddEndPuncttrue
\mciteSetBstMidEndSepPunct{\mcitedefaultmidpunct}
{\mcitedefaultendpunct}{\mcitedefaultseppunct}\relax
\EndOfBibitem
\bibitem[Luo \latin{et~al.}(2023)Luo, Li, Wang, Guo, Lin, Zhao, Hu, Zhu, Xu, Li, and Guo]{Luofabrication2023}
Luo,~Q.-Y.; Li,~Q.; Wang,~J.-F.; Guo,~P.-J.; Lin,~W.-X.; Zhao,~S.; Hu,~Q.-C.; Zhu,~Z.-Q.; Xu,~J.-S.; Li,~C.-F.; Guo,~G.-C. Fabrication and quantum sensing of spin defects in silicon carbide. \emph{Frontiers in Physics} \textbf{2023}, \emph{11}, 1--16\relax
\mciteBstWouldAddEndPuncttrue
\mciteSetBstMidEndSepPunct{\mcitedefaultmidpunct}
{\mcitedefaultendpunct}{\mcitedefaultseppunct}\relax
\EndOfBibitem
\bibitem[He \latin{et~al.}(2023)He, Li, Wen, Zhou, Lin, Hao, Xu, Li, and Guo]{Zhen-Xuan2023}
He,~Z.-X.; Li,~Q.; Wen,~X.-L.; Zhou,~J.-Y.; Lin,~W.-X.; Hao,~Z.-H.; Xu,~J.-S.; Li,~C.-F.; Guo,~G.-C. Maskless Generation of Single Silicon Vacancy Arrays in Silicon Carbide by a Focused He+ Ion Beam. \emph{ACS Photonics} \textbf{2023}, \emph{10}, 2234--2240\relax
\mciteBstWouldAddEndPuncttrue
\mciteSetBstMidEndSepPunct{\mcitedefaultmidpunct}
{\mcitedefaultendpunct}{\mcitedefaultseppunct}\relax
\EndOfBibitem
\bibitem[Brereton \latin{et~al.}(2020)Brereton, Puent, Vanhoy, Glaser, and Carter]{BRERETON2020114014}
Brereton,~P.; Puent,~D.; Vanhoy,~J.; Glaser,~E.; Carter,~S. Spin coherence as a function of depth for high-density ensembles of silicon vacancies in proton-irradiated 4H–SiC. \emph{Solid State Communications} \textbf{2020}, \emph{320}, 114014\relax
\mciteBstWouldAddEndPuncttrue
\mciteSetBstMidEndSepPunct{\mcitedefaultmidpunct}
{\mcitedefaultendpunct}{\mcitedefaultseppunct}\relax
\EndOfBibitem
\bibitem[Kraus \latin{et~al.}(2017)Kraus, Simin, Kasper, Suda, Kawabata, Kada, Honda, Hijikata, Ohshima, Dyakonov, \latin{et~al.} others]{kraus2017three}
Kraus,~H.; Simin,~D.; Kasper,~C.; Suda,~Y.; Kawabata,~S.; Kada,~W.; Honda,~T.; Hijikata,~Y.; Ohshima,~T.; Dyakonov,~V.; others Three-dimensional proton beam writing of optically active coherent vacancy spins in silicon carbide. \emph{Nano letters} \textbf{2017}, \emph{17}, 2865--2870\relax
\mciteBstWouldAddEndPuncttrue
\mciteSetBstMidEndSepPunct{\mcitedefaultmidpunct}
{\mcitedefaultendpunct}{\mcitedefaultseppunct}\relax
\EndOfBibitem
\bibitem[Ohshima \latin{et~al.}(2018)Ohshima, Satoh, Kraus, Astakhov, Dyakonov, and Baranov]{ohshima2018creation}
Ohshima,~T.; Satoh,~T.; Kraus,~H.; Astakhov,~G.~V.; Dyakonov,~V.; Baranov,~P.~G. Creation of silicon vacancy in silicon carbide by proton beam writing toward quantum sensing applications. \emph{Journal of Physics D: Applied Physics} \textbf{2018}, \emph{51}, 333002\relax
\mciteBstWouldAddEndPuncttrue
\mciteSetBstMidEndSepPunct{\mcitedefaultmidpunct}
{\mcitedefaultendpunct}{\mcitedefaultseppunct}\relax
\EndOfBibitem
\bibitem[Chen \latin{et~al.}(2019)Chen, Salter, Niethammer, Widmann, Kaiser, Nagy, Morioka, Babin, Erlekampf, Berwian, \latin{et~al.} others]{chen2019laser}
Chen,~Y.-C.; Salter,~P.~S.; Niethammer,~M.; Widmann,~M.; Kaiser,~F.; Nagy,~R.; Morioka,~N.; Babin,~C.; Erlekampf,~J.; Berwian,~P.; others Laser writing of scalable single color centers in silicon carbide. \emph{Nano Lett.} \textbf{2019}, \emph{19}, 2377--2383\relax
\mciteBstWouldAddEndPuncttrue
\mciteSetBstMidEndSepPunct{\mcitedefaultmidpunct}
{\mcitedefaultendpunct}{\mcitedefaultseppunct}\relax
\EndOfBibitem
\bibitem[Day \latin{et~al.}(2023)Day, Dietz, Sutula, Yeh, and Hu]{day2023laser}
Day,~A.~M.; Dietz,~J.~R.; Sutula,~M.; Yeh,~M.; Hu,~E.~L. Laser writing of spin defects in nanophotonic cavities. \emph{Nature Materials} \textbf{2023}, \emph{22}, 696--702\relax
\mciteBstWouldAddEndPuncttrue
\mciteSetBstMidEndSepPunct{\mcitedefaultmidpunct}
{\mcitedefaultendpunct}{\mcitedefaultseppunct}\relax
\EndOfBibitem
\bibitem[Castelletto \latin{et~al.}(2018)Castelletto, Almutairi, Kumagai, Katkus, Hayasaki, Johnson, and Juodkazis]{castelletto2018photoluminescence}
Castelletto,~S.; Almutairi,~A. F.~M.; Kumagai,~K.; Katkus,~T.; Hayasaki,~Y.; Johnson,~B.; Juodkazis,~S. Photoluminescence in hexagonal silicon carbide by direct femtosecond laser writing. \emph{Optics letters} \textbf{2018}, \emph{43}, 6077--6080\relax
\mciteBstWouldAddEndPuncttrue
\mciteSetBstMidEndSepPunct{\mcitedefaultmidpunct}
{\mcitedefaultendpunct}{\mcitedefaultseppunct}\relax
\EndOfBibitem
\bibitem[Castelletto \latin{et~al.}(2021)Castelletto, Maksimovic, Katkus, Ohshima, Johnson, and Juodkazis]{castelletto2021color}
Castelletto,~S.; Maksimovic,~J.; Katkus,~T.; Ohshima,~T.; Johnson,~B.~C.; Juodkazis,~S. Color centers enabled by direct femto-second laser writing in wide bandgap semiconductors. \emph{Nanomaterials} \textbf{2021}, \emph{11}, 72\relax
\mciteBstWouldAddEndPuncttrue
\mciteSetBstMidEndSepPunct{\mcitedefaultmidpunct}
{\mcitedefaultendpunct}{\mcitedefaultseppunct}\relax
\EndOfBibitem
\bibitem[Almutairi \latin{et~al.}(2022)Almutairi, Partridge, Xu, Cole, and Holland]{almutairi2022direct}
Almutairi,~A. F.~M.; Partridge,~J.~G.; Xu,~C.; Cole,~I.~S.; Holland,~A.~S. {Direct writing of divacancy centers in silicon carbide by femtosecond laser irradiation and subsequent thermal annealing}. \emph{Applied Physics Letters} \textbf{2022}, \emph{120}, 014003\relax
\mciteBstWouldAddEndPuncttrue
\mciteSetBstMidEndSepPunct{\mcitedefaultmidpunct}
{\mcitedefaultendpunct}{\mcitedefaultseppunct}\relax
\EndOfBibitem
\bibitem[Wang \latin{et~al.}(2020)Wang, Liu, Yan, Li, Yang, Guo, Zhou, Huang, Xu, Li, and Guo]{Wangexperimental2020}
Wang,~J.-F.; Liu,~Z.-H.; Yan,~F.-F.; Li,~Q.; Yang,~X.-G.; Guo,~L.; Zhou,~X.; Huang,~W.; Xu,~J.-S.; Li,~C.-F.; Guo,~G.-C. Experimental Optical Properties of Single Nitrogen Vacancy Centers in Silicon Carbide at Room Temperature. \emph{ACS Photonics} \textbf{2020}, \emph{7}, 1611--1616\relax
\mciteBstWouldAddEndPuncttrue
\mciteSetBstMidEndSepPunct{\mcitedefaultmidpunct}
{\mcitedefaultendpunct}{\mcitedefaultseppunct}\relax
\EndOfBibitem
\bibitem[An \latin{et~al.}(2022)An, Wang, and Fang]{10119524}
An,~H.; Wang,~J.; Fang,~F. Surface modification of silicon carbide at atomic and close-to-atomic scale by femtosecond laser. 2022 8th International Conference on Nanomanufacturing and 4th AET Symposium on ACSM and Digital Manufacturing (Nanoman-AETS). 2022; pp 1--5\relax
\mciteBstWouldAddEndPuncttrue
\mciteSetBstMidEndSepPunct{\mcitedefaultmidpunct}
{\mcitedefaultendpunct}{\mcitedefaultseppunct}\relax
\EndOfBibitem
\bibitem[Wang \latin{et~al.}(2024)Wang, Fang, Li, Wang, and Sun]{wang2024laser}
Wang,~X.-J.; Fang,~H.-H.; Li,~Z.-Z.; Wang,~D.; Sun,~H.-B. Laser manufacturing of spatial resolution approaching quantum limit. \emph{Light: Science \& Applications} \textbf{2024}, \emph{13}, 6\relax
\mciteBstWouldAddEndPuncttrue
\mciteSetBstMidEndSepPunct{\mcitedefaultmidpunct}
{\mcitedefaultendpunct}{\mcitedefaultseppunct}\relax
\EndOfBibitem
\bibitem[Kucsko \latin{et~al.}(2013)Kucsko, Maurer, Yao, Kubo, Noh, Lo, Park, and Lukin]{kucsko2013nanometre}
Kucsko,~G.; Maurer,~P.~C.; Yao,~N.~Y.; Kubo,~M.; Noh,~H.~J.; Lo,~P.~K.; Park,~H.; Lukin,~M.~D. Nanometre-scale thermometry in a living cell. \emph{Nature} \textbf{2013}, \emph{500}, 54--58\relax
\mciteBstWouldAddEndPuncttrue
\mciteSetBstMidEndSepPunct{\mcitedefaultmidpunct}
{\mcitedefaultendpunct}{\mcitedefaultseppunct}\relax
\EndOfBibitem
\end{mcitethebibliography}
\end{document}